\documentclass[11pt, a4paper]{article}

\usepackage{amsmath,amssymb,mathrsfs}
\usepackage{subfigure}
\usepackage[dvipdfmx]{graphicx}
\usepackage{cite}

\usepackage{url}

\begin{document}

\renewcommand{\theequation}{\thesection.\arabic{equation}}

\renewcommand{\thefootnote}{\fnsymbol{footnote}}
\setcounter{footnote}{0}

\def\thefootnote{\fnsymbol{footnote}}


{\hfill DESY 17-036}

\vspace{4.5cm}

\begin{center}

{\textbf{\Large
A review of gravitational waves from cosmic domain walls
}}

\bigskip

\vspace{0.4 truecm}

{\bf Ken'ichi Saikawa} \\[5mm]

\begin{tabular}{c}
{\em Deutsches Elektronen-Synchrotron DESY,}\\
{\em Notkestrasse 85, 22607 Hamburg, Germany}\\[.4em]
\end{tabular}

\vspace{3cm}

{\bf Abstract}
\end{center}

\begin{quote}
In this contribution, we discuss the cosmological scenario where unstable domain walls are formed in the early universe and their late-time annihilation 
produces a significant amount of gravitational waves. After describing cosmological constraints on long-lived domain walls, we estimate the typical amplitude 
and frequency of gravitational waves observed today. We also review possible extensions of the standard model of particle physics that predict the formation of 
unstable domain walls and can be probed by observation of relic gravitational waves. It is shown that recent results of pulser timing arrays and direct detection 
experiments partially exclude the relevant parameter space, and that a much wider parameter space can be covered by
the next generation of gravitational wave observatories.
\end{quote}

\thispagestyle{empty}
\vfill

\renewcommand{\thepage}{\arabic{page}}
\setcounter{page}{1}

\newpage

\tableofcontents

\renewcommand{\thefootnote}{\arabic{footnote}}
\setcounter{footnote}{0}

\section{Introduction}
\label{sec1}

The progress of direct observations~\cite{Abbott:2016blz,Abbott:2016nmj} of gravitational waves (GWs) is bringing
about drastic developments in astrophysics and cosmology.
We expect to obtain a lot of important information about physics at very high energies from direct observations of GWs
due to the fact that they interact very weakly with matter and hence preserve almost all the
features characterizing astrophysical or cosmological events~\cite{Maggiore:1900zz,Maggiore:1999vm}.
Experimental sensitivities for the direct detection of GWs have been improved substantially during the past decades,
and many new GW observatories are now planned to be built in the world.
In this context, it is worth investigating various possible sources of GWs and
clarifying to what extent we can extract the information about new physics from future observations.
So far, various cosmological sources of relic GWs are discussed in the literature, such as
the primordial amplification of vacuum fluctuations~\cite{Grishchuk:1974ny,Starobinsky:1979ty,Smith:2005mm,Zhao:2009pt,Corda:2009bx,Zhao:2010ic}, 
cosmological phase transitions~\cite{Witten:1984rs,Kamionkowski:1993fg}, 
cosmic strings~\cite{Vilenkin:1981bx,Accetta:1988bg,Caldwell:1991jj},
and preheating after inflation~\cite{Khlebnikov:1997di,Easther:2006gt,GarciaBellido:2007af,Dufaux:2008dn,Kawasaki:2012rw}.
Furthermore, the recent starting of the GW astronomy can provide a distinctive way of testing General Relativity and other theories of gravity~\cite{Corda:2009re}.

In this article, we consider domain walls as possible cosmological sources of GWs.
Domain walls are sheet-like topological defects, which might be created in the early universe when a discrete symmetry is spontaneously broken~\cite{Kibble:1976sj}.
Since discrete symmetries are ubiquitous in high energy physics beyond the Standard Model (SM),
many new physics models predict the formation of domain walls in the early universe.
By considering their cosmological evolution,
it is possible to deduce several constraints on such models
even if their energy scales are much higher than that probed in the laboratory experiments.

In general, the formation of domain walls is regarded as a problem in cosmology~\cite{Zeldovich:1974uw},
since their energy density soon dominates the total energy density of the universe, which conflicts with the present observational results.
However, we can consider the possibility that domain walls are unstable and collapse before they overclose the universe~\cite{Vilenkin:1981zs,Gelmini:1988sf,Larsson:1996sp}.
Their unstability might be guaranteed if the discrete symmetry is only approximate and explicitly broken by a small parameter in the theory.
In such a scenario, a significant amount of GWs can be produced during the process of collisions and annihilations of domain walls,
and they may remain as a stochastic GW background in the present universe.
Observations of such relic GWs will enable us to trace the events in the very early universe
and provide a new way of investigating physics at very high energies.

The purpose of this article is to review the physics of cosmic domain walls and to evaluate the detectability of
GWs produced by them in present and future observations.
In particular, we summarize the results of
recent theoretical developments including various particle physics motivations so far proposed in the literature and
the methods to estimate the GW signatures based on field theoretic lattice simulations.

We note that the production mechanism of GWs discussed in this article is different from that discussed in 
the context of phase transitions [{\it e.g.} Refs.~\cite{Witten:1984rs,Kamionkowski:1993fg,Grojean:2006bp,Dev:2016feu,Balazs:2016tbi,Ivanov:2017dad}].
In the latter case, a strong first order phase transition is assumed, and GWs are produced due to the collision of bubbles and subsequent turbulences.
On the other hand, in the former case it is not necessary to assume the first order phase transition,
and GWs are produced due to the late-time motion and interaction of domain walls.
The key ingredient of this scenario is the existence of quasi-degenerate vacua after the phase transition.

This review is organized as follows.
The theoretical basics of domain walls and cosmological constraints on them are described in Sec.~\ref{sec2}.
The semi-analytical approach to estimate the production of GWs from domain walls is discussed in Sec.~\ref{sec3}.
Section~\ref{sec4} deals with particle physics models that predict the production of a significant amount of GWs from domain walls.
In Sec.~\ref{sec5}, we compare the GW signatures with sensitivities of present and future experiments.
Finally, we conclude in Sec.~\ref{sec6}.

\section{Domain walls and cosmology}
\label{sec2}
\setcounter{equation}{0}

\subsection{Field theory}
\label{sec:field_theory}

As an illustrative example, let us consider the following toy model of a real scalar field $\phi$,
\begin{align}
\mathcal{L} &= -\frac{1}{2}\partial^{\mu}\phi\partial_{\mu}\phi - V(\phi), \label{Lagrangian_Z2}\\
V(\phi) &= \frac{\lambda}{4}\left(\phi^2-v^2\right)^2. \label{potential_Z2}
\end{align}
Note that the potential $V(\phi)$ has two degenerate minima at $\phi=\pm v$.
In this theory there is a discrete $Z_2$ symmetry, under which the field transforms as $\phi\to -\phi$.
This discrete symmetry is spontaneously broken when the scalar field acquires a vacuum expectation value (VEV), $\langle\phi\rangle = \pm v$.
The scalar field takes one of the two discrete values ($+v$ and $-v$) after the spontaneous symmetry breaking,
which means that two different domains can appear.
Domain walls are produced around the boundary of these two domains.

Consider a static planar domain wall configuration lying perpendicular to the $z$-axis in the Minkowski space, $\phi=\phi(z)$.
Solving the field equation $d^2\phi/dz^2 - dV/d\phi = 0$, we obtain
\begin{equation}
\phi(z) = v\tanh\left[\sqrt{\frac{\lambda}{2}}vz\right].
\end{equation}
We see that $\phi(z)$ approaches $\pm v$ as $z\to \pm\infty$, and it rapidly changes around $z=0$. 
The width of the domain wall $\delta$ can be estimated as a typical length scale of the spatial variation of $\phi(z)$,
\begin{equation}
\delta \simeq \left(\sqrt{\frac{\lambda}{2}}v\right)^{-1}. \label{width_Z2}
\end{equation}

The energy-momentum tensor for the static solution $\phi=\phi(z)$ is given by
\begin{align}
T_{\mu\nu}(z) = \left(\frac{d\phi(z)}{dz}\right)^2\mathrm{diag}(+1,-1,-1,0).
\end{align} 
Integrating $T_{00}$ over the direction perpendicular to the wall, we obtain its surface energy density,
\begin{equation}
\sigma = \int^{\infty}_{-\infty} dzT_{00} = \frac{4}{3}\sqrt{\frac{\lambda}{2}} v^3. \label{tension_Z2}
\end{equation}
Note that the integration of the spatial components results in the same quantity:
$\int dz T_{11} = \int dz T_{22} = -\sigma$.
Therefore, $\sigma$ is also referred to as the \emph{tension} of domain walls.

Another interesting example is the following model of a real scalar field $a$,
\begin{equation}
\mathcal{L} = -\frac{1}{2}\partial_{\mu}a\partial^{\mu}a - \frac{m^2 v^2}{N^2}\left[1-\cos\left(N\frac{a}{v}\right)\right], \label{Lagrangian_ZN}
\end{equation}
where the field $a$ is defined within a finite domain $[0,2\pi v]$, and $N$ is a positive integer.
This kind of potential naturally arises in the context of axion models, which will be discussed in Sec.~\ref{sec:axion}.
In this model, there is a discrete $Z_N$ symmetry under which the field transforms as $a/v \to a/v + 2\pi k/N$ with $k=0,1,\dots, N-1$. 
This symmetry is spontaneously broken once the field $a$ settles down to one of $N$ degenerate minima of the potential,
and domain walls are formed around the boundary of these vacua.

If we consider a planar wall orthogonal to $z$-axis $a=a(z)$, the solution of the classical field equation reads
\begin{equation}
\frac{a(z)}{v} = \frac{2\pi k}{N} + \frac{4}{N}\tan^{-1}\exp\left(mz\right).
\end{equation}
This configuration interpolates between two vacua, $a/v = 2\pi k/N$ at $z\to -\infty$ and $a/v = 2\pi(k+1)/N$ at $z\to +\infty$.
From the above solution, we can estimate the thickness of the wall as
\begin{equation}
\delta \simeq m^{-1}. \label{width_ZN}
\end{equation}
The tension of domain walls reads
\begin{equation}
\sigma = \int^{\infty}_{-\infty} dz \left(\frac{da}{dz}\right)^2 = \frac{8mv^2}{N^2}. \label{tension_ZN}
\end{equation}

As we have seen in above two examples, the properties of domain walls can be characterized by two model-dependent quantities,
the tension $\sigma$ and the thickness $\delta$.
In general, the wall thickness is roughly given by Compton wavelength of the field which causes the spontaneous breaking of discrete symmetry,
while the tension is estimated in terms of the height of the potential energy $V_0$ separating the degenerate minima,
\begin{equation}
\sigma \sim \delta \cdot V_0. \label{typical_estimate_sigma}
\end{equation}
In the following, we do not specify the magnitude of $\sigma$ and provide some model-independent arguments 
about the evolution of domain walls. We will come back to the mode-dependent issues in Sec.~\ref{sec4}.

\subsection{Cosmological evolution}

Domain walls exist if different vacua are populated in the universe.
Whether such a distribution of vacua appears or not depends on the cosmological initial conditions.
In particular, it is widely believed that the universe underwent a period of exponentially rapid expansion, called inflation.
Depending on the conditions at the inflationary epoch, the formation of domain walls must be seriously taken into account.

Suppose that the toy model scalar field $\phi$ introduced in the previous subsection stayed at a certain vacuum 
before the inflationary period.
In such a setup, we naively expect that domain walls do not exist in the present universe, since a domain on which $\langle\phi\rangle$
takes an uniform value exponentially glows during inflation and the size of such a domain is much larger than the present horizon size.
However, such a naive expectation is not necessarily true.
During inflation, the field $\phi$ acquires vacuum fluctuations of order $\delta\phi \sim H_{\rm inf}/2\pi$
if its effective mass $m_{\phi}$ [$m_{\phi}^2=2\lambda v^2$ in the model given by Eq.~\eqref{potential_Z2}]
is smaller than $H_{\rm inf}$~\cite{Vilenkin:1982wt,Linde:1982uu,Starobinsky:1982ee},
where $H_{\rm inf}$ is the Hubble parameter during inflation.
Once such a condition is satisfied, the $\phi$ field easily jumps into other domains within one Hubble time,
and as a result many different domains can exist after inflation, which leads to the formation of domain walls.
Furthermore, even if $H_{\rm inf}$ is sufficiently small such that the $\phi$ field never acquires large fluctuations during inflation,
it can thermalize and have thermal fluctuations due to the reheating after inflation.
If this is the case, the discrete symmetry is thermally restored when the maximum temperature after inflation $T_{\rm max}$
becomes larger than $m_{\phi}$. After that, domain walls are formed when the universe cools below some critical temperature.
Therefore, we expect that the formation of domain walls can happen if either the Hubble parameter during inflation $H_{\rm inf}$
or the maximum temperature after inflation $T_{\rm max}$ is sufficiently larger than the mass $m_{\phi}$ of the field $\phi$.\footnote{We emphasize that
the condition is not robust, and we can consider several loopholes depending on the details of the models.
For instance, if the $\phi$ field never thermalizes, domain walls may not be formed even when $T_{\rm max}> m_{\phi}$ is satisfied. 
We can also consider the case where the effective mass $m_{\phi,{\rm eff}}$ during inflation is different from the bare mass $m_{\phi}$.
In such a case, it is possible to avoid the formation of domain walls even when $H_{\rm inf} > m_{\phi}$ is satisfied~\cite{Harigaya:2015hha,Mazumdar:2015dwd}.}

After the formation of domain walls, their dynamics can be described by two kinds of forces.
One is the tension force, which is given by
\begin{equation}
p_T \sim \frac{\sigma}{R_{\rm wall}},
\end{equation}
where $R_{\rm wall}$ represents a typical curvature radius of the walls.
The other is the friction force, which appears if there is a interaction of the field composing the core of domain walls and particles in thermal bath.
It can be estimated as~\cite{Vilenkin:2000jqa}
\begin{equation}
p_F \sim \Delta p \cdot n \sim v T^4,
\end{equation}
where $\Delta p\sim Tv$ is a typical momentum transfer due to the collision with a particle,
$n\sim T^3$ is the number density of particles, and $v$ is the velocity of domain walls.
These two forces are balanced, $p_T \sim p_F$, and from this condition we obtain 
\begin{align}
v \sim \frac{\sigma}{T^4 R_{\rm wall}} \sim \frac{(\sigma t)^{1/2}}{M_{\rm Pl}}, \label{v_friction_regime}\\
R_{\rm wall} \sim vt \sim \frac{\sigma^{1/2}t^{3/2}}{M_{\rm Pl}}, \label{Rwall_friction_regime}
\end{align}
where we have used $T^4 \sim M_{\rm Pl}^2/t^2$ assuming the radiation dominated background, and
$M_{\rm Pl} \simeq 2.435\times 10^{18}\,\mathrm{GeV}$ is the reduced Planck mass.
Let $t_r$ denote the time at which domain walls become relativistic.
Equations~\eqref{v_friction_regime} and~\eqref{Rwall_friction_regime} imply that
their curvature radius becomes comparable to the horizon size at that time, $R_{\rm wall} \sim M_{\rm Pl}^2/\sigma \sim t_r$.
We also see that their energy density $\rho_{\rm wall}$ dominates over the total energy density of the universe $\rho_c$ at that time,
\begin{equation} 
\rho_{\rm wall} \sim \frac{\sigma}{R_{\rm wall}} \sim \frac{\sigma^2}{M_{\rm Pl}^2} \sim \frac{M_{\rm Pl}^2}{t_r^2} \sim \rho_c(t_r).
\end{equation}
In other words, they remain non-relativistic as long as their energy density is subdominant.

The friction force is exponentially damped when the temperature of the background radiations becomes less than the mass of particles that interact with
domain walls. Therefore, we expect that the effect of the friction force becomes negligible at sufficiently late times
if domain walls only interact with massive particle states. On the other hand, if they interact with lighter particles such as those in the SM,
we must carefully evaluate the effect of the friction force in order to describe their late time evolution.
In what follows, we focus on the case where the friction force becomes negligible at sufficiently early times.

Once the friction force becomes irrelevant, the dynamics of domain walls is dominated by the tension force, which stretches them up to the horizon size.
Many numerical studies~\cite{Press:1989yh,Garagounis:2002kt,Oliveira:2004he,Avelino:2005kn,Leite:2011sc,Leite:2012vn,Martins:2016ois}
confirmed that the evolution of domain walls in this regime can be described by the {\it scaling solution}, in which
their energy density evolves according to the simple scaling law $\rho_{\rm wall} \propto t^{-1}$, and their typical size is given by the Hubble radius $\sim t$.\footnote{We note that
the results of numerical simulations imply $\rho_{\rm wall} \propto t^{-\nu}$, where the exponent $\nu$ slightly deviates from $\nu=1$.
At this point it is unclear whether this deviation represents some physical effect or just a numerical artifact which could be removed if we improve the dynamical range of the simulation.
In this article, we carry out the analysis by assuming that the evolution of domain walls is described by the exact scaling law [Eq.~\eqref{rho_wall_scaling}].}
An analytic method to calculate the evolution of domain walls was also proposed in Refs.~\cite{Hindmarsh:1996xv,Hindmarsh:2002bq}, which again showed the existence of the scaling solution.

It will be convenient to parameterize the energy density of domain walls in the scaling regime as
\begin{equation}
\rho_{\rm wall}(t) = \mathcal{A}\frac{\sigma}{t}, \label{rho_wall_scaling}
\end{equation}
where $\mathcal{A}$ is a parameter which takes an almost constant value during the scaling regime, and we call it an area parameter.
According to the results of field theoretic simulations of domain walls in the $Z_2$ symmetric model [Eq.~\eqref{potential_Z2}] performed in Ref.~\cite{Hiramatsu:2013qaa}, we have
\begin{equation}
\mathcal{A} \simeq 0.8\pm0.1, \label{area_parameter_value}
\end{equation}
where the error corresponds to the statistical uncertainty caused by different realizations of initial conditions for the simulations.
The area parameter was also estimated for domain walls in the $Z_N$ symmetric model [Eq.~\eqref{Lagrangian_ZN}] in Refs.~\cite{Hiramatsu:2012sc,Kawasaki:2014sqa}.
It was shown that the value of $\mathcal{A}$ for the case with $N=2$ agrees with Eq.~\eqref{area_parameter_value}, and that it increases proportionally with $N$.

The energy density of domain walls in the scaling regime $\rho_{\rm wall} \propto t^{-1}$
decays slower than that of cold matters $\rho_{\rm matter} \propto R^{-3}(t)$ and radiations $\rho_{\rm rad} \propto R^{-4}(t)$,
where $R(t)$ is the scale factor of the universe.
Therefore, they gradually dominate the energy density of the universe.
From the condition $\rho_c(t) = \rho_{\rm wall}(t)$, we estimate the time at which the wall domination occurs,
\begin{align}
t_{\rm dom} &= \frac{3M_{\rm Pl}^2}{4\mathcal{A}\sigma} \nonumber\\
&\simeq 2.93 \times 10^3\,\mathrm{sec}\,\mathcal{A}^{-1}\left(\frac{\sigma}{\mathrm{TeV}^3}\right)^{-1}. \label{t_dom}
\end{align}
Here we assumed that the total energy density of the universe is dominated by radiations before $t=t_{\rm dom}$, {\it i.e.}, $\rho_c = 3M_{\rm Pl}^2/4t^2$.
Once domain walls dominate the energy density of the universe, the subsequent evolution of the universe is drastically altered.
The equation of state for an isotropic gas of non-relativistic domain walls is given by $w=-2/3$~\cite{Vilenkin:2000jqa},
which implies that the scale factor in the wall dominated universe evolves as
\begin{equation}
R(t) \propto t^2.
\end{equation}
Such a rapid expansion is incompatible with standard cosmology.

Even if the energy density of domain walls is subdominant at the present time, they may cause another problem.
Since their typical curvature radius is comparable to the Hubble radius, they introduce large scale density fluctuations,
whose magnitude is estimated as
\begin{equation}
\frac{\delta \rho}{\rho} \sim \frac{\rho_{\rm wall}}{\rho_c} \sim G\sigma t_0 \sim 10^{12}\left(\frac{\sigma}{\mathrm{TeV}^3}\right),
\end{equation}
where $G$ is Newton's gravitational constant, and we used $t_0\sim H^{-1}_0$ with $t_0$ and $H_0$ being the present cosmic time and the Hubble constant, respectively.
The observation of the cosmic microwave background radiation implies $\delta\rho/\rho \lesssim \mathcal{O}(10^{-5})$,
from which we obtain the following condition
\begin{equation}
\sigma^{1/3} \lesssim \mathcal{O}(\mathrm{MeV}). \label{ZKO_bound}
\end{equation}
This constraint was first discussed in Ref.~\cite{Zeldovich:1974uw}, and it is referred to as the Zel'dovich-Kobzarev-Okun bound.
We see that domain walls with a tension as large as $\sigma > \mathcal{O}(\mathrm{MeV}^3)$ must not exist in the universe at the present time.

\subsection{Biased domain walls}

\begin{figure}[htbp]
\begin{center}
\includegraphics[scale=0.8]{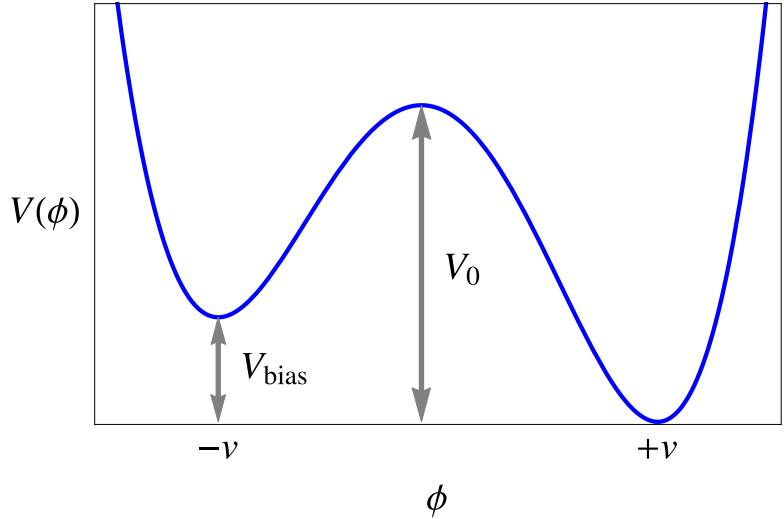}
\caption{Sketch of the biased potential given by Eqs.~\eqref{potential_Z2} and~\eqref{bias_potential_Z2}.}
\label{fig:Z2_biased_potential}
\end{center}
\end{figure}

One possible solution to the domain wall problem is to introduce an energy bias in the potential, which lifts the degenerate minima~\cite{Vilenkin:1981zs,Gelmini:1988sf,Larsson:1996sp}.\footnote{It is also possible to
avoid the domain wall problem by assuming an asymmetric probability distribution for initial field fluctuations~\cite{Coulson:1995nv} instead of introducing the energy bias in the potential.
Here we do not consider such a scenario, since it depends on the models of the evolution of the early universe, which must produce an appropriate initial field distribution.}
Let us consider the model for the real scalar field $\phi$
discussed in Sec.~\ref{sec:field_theory}. Here we artificially introduce the following term
\begin{equation}
\Delta V(\phi) = \epsilon v\phi\left(\frac{1}{3}\phi^2 - v^2\right), \label{bias_potential_Z2}
\end{equation}
in addition to Eq.~\eqref{potential_Z2}, where $\epsilon$ is a dimensionless constant. The modified potential is shown in Figure~\ref{fig:Z2_biased_potential}.
This potential has minima at $\phi = \pm v$, but there is an energy difference between them, 
\begin{equation}
V_{\rm bias} \equiv V(-v) - V(+v) = \frac{4}{3}\epsilon v^4. \label{energy_bias_Z2}
\end{equation}
Because of the existence of this energy difference, domain walls become unstable and eventually collapse.
Note that the additional term~\eqref{bias_potential_Z2} explicitly breaks the discrete $Z_2$ symmetry.
Therefore, this solution works if the discrete symmetry is not exact, but holds only approximately.

We note that domain walls cannot be created from the beginning
if the energy deference between two vacua $V_{\rm bias}$ is sufficiently large~\cite{Gelmini:1988sf}.
In order to clarify the condition to have domain walls in the presence of the energy bias,
let us consider the probabilities $p_+$ and $p_-$ in which the scalar field ends up in the plus vacuum ($\phi=+v$)
and the minus vacuum ($\phi=-v$), respectively, after the phase transition.
The ratio between these two probabilities is given by
\begin{equation}
\frac{p_-}{p_+} = \exp\left(-\frac{\Delta F}{T}\right) \simeq \exp\left(-\frac{V_{\rm bias}}{V_0}\right),
\end{equation}
where $\Delta F = V_{\rm bias}\cdot \xi^3$ is the difference of the free energy between two vacua,
$\xi$ is the correlation length at the epoch of the phase transition, and we estimate $T$ as the Ginzburg temperature, $T \simeq V_0\cdot \xi^3$,
with $V_0$ being the height of the potential barrier between two minima.
The above equation implies that the spatial distribution of two vacua after the phase transition becomes asymmetric if $V_{\rm bias} \ne 0$.
According to the prediction of percolation theory, the critical value above which an infinite cluster of the minus vacuum appears in the space
is given by $p_c=0.311$, if the system is treated as a three dimensional cubic lattice~\cite{Stauffer:1978kr}. Requiring that a large cluster of the false vacuum
appears in the space ($p_- > p_c$), we obtain
\begin{equation}
\frac{V_{\rm bias}}{V_0} < \ln\left(\frac{1-p_c}{p_c}\right) = 0.795. \label{bound_bias}
\end{equation}
In other words, large scale domain walls are expected to be formed as long as the above condition is satisfied.

Even if $V_{\rm bias}$ is sufficiently small such that domain walls are created at the phase transition,
the false vacuum region tends to shrink due to the existence of the energy difference:
There is a volume pressure force acting on the walls, whose magnitude is estimated as $p_V \sim V_{\rm bias}$.
The collapse of domain walls happens when this pressure force becomes greater than the tension force $p_T\sim \sigma/R_{\rm wall}$.
If we assume that domain walls have reached the scaling regime beforehand, their typical curvature radius is given by $R_{\rm wall}\simeq t/\mathcal{A}$ [see Eq.~\eqref{rho_wall_scaling}].
Hence, from the condition that two forces become comparable, $p_V \sim p_T$, we can estimate their annihilation time:
\begin{align}
t_{\rm ann} &= C_{\rm ann} \frac{\mathcal{A}\sigma}{V_{\rm bias}} \nonumber\\
& = 6.58\times 10^{-4}\,\mathrm{sec}\, C_{\rm ann}\mathcal{A}\left(\frac{\sigma}{\mathrm{TeV}^3}\right)\left(\frac{V_{\rm bias}}{\mathrm{MeV}^4}\right)^{-1}, \label{t_ann_general}
\end{align}
where $C_{\rm ann}$ is a coefficient of $\mathcal{O}(1)$.
If the annihilation occurs in the radiation dominated era, the temperature at $t=t_{\rm ann}$ is given by
\begin{equation}
T_{\rm ann} = 3.41 \times 10^{-2}\,\mathrm{GeV}\,C_{\rm ann}^{-1/2}\mathcal{A}^{-1/2}\left(\frac{g_*(T_{\rm ann})}{10}\right)^{-1/4}\left(\frac{\sigma}{\mathrm{TeV}^3}\right)^{-1/2}\left(\frac{V_{\rm bias}}{\mathrm{MeV}^4}\right)^{1/2}, \label{T_ann_general}
\end{equation}
where $g_*(T)$ is the relativistic degrees of freedom for the radiation energy density at a given temperature $T$.
The value of $C_{\rm ann}$ (or $C_d$ in Ref.~\cite{Kawasaki:2014sqa}) can be determined from numerical simulations.
It typically takes the value of $C_{\rm ann}\simeq 2\textendash 5$, depending on $N$ for the model with the $Z_N$ symmetry [Eq.~\eqref{Lagrangian_ZN}].
It also depends on the choice of the criterion to determine the decay time of domain walls in the simulations. For more details, see Ref.~\cite{Kawasaki:2014sqa}.

Note that the lifetime $t_{\rm ann}$ is inversely proportional to $V_{\rm bias}$.
If the energy bias is sufficiently small, domain walls live for a long time.
Requiring that their collapse occurs before they overclose the universe $t_{\rm ann} < t_{\rm dom}$ [see Eq.~\eqref{t_dom}],
we obtain the lower bound\footnote{The domain wall domination
does not directly imply a cosmological disaster. In principle, it can happen in the early universe without causing any trouble with the standard cosmology if such domain walls are annihilated before the epoch of Big Bang nucleosynthesis (BBN). 
For instance, the possibilities of diluting unwanted relics in the domain wall dominated universe are discussed in Refs.~\cite{Kawasaki:2004rx,Hattori:2015xla}.
However, little is known about the detailed dynamics of domain walls in the domain wall dominated universe, and their behavior in such a scenario is uncertain. Therefore, in this work we just focus on the case in which the energy density of domain walls never dominates the critical energy density of the universe, and use the condition of the domain wall domination to indicate the potential uncertainties.}
on the magnitude of the energy bias, $V_{\rm bias}>4 C_{\rm ann}\mathcal{A}^2\sigma^2/3 M_{\rm Pl}^2$, or
\begin{equation}
V_{\rm bias}^{1/4} > 2.18 \times 10^{-5}\,\mathrm{GeV}\,C_{\rm ann}^{1/4}\mathcal{A}^{1/2}\left(\frac{\sigma}{\mathrm{TeV}^3}\right)^{1/2}.
\end{equation}
By using Eq.~\eqref{T_ann_general}, this condition can be rewritten in terms of the annihilation temperature,
\begin{equation}
T_{\rm ann} > 1.62\times 10^{-5}\,\mathrm{GeV}\,\mathcal{A}^{1/2}\left(\frac{g_*(T_{\rm ann})}{10}\right)^{-1/4}\left(\frac{\sigma}{\mathrm{TeV}^3}\right)^{1/2}. \label{wall_domination_constraint}
\end{equation}

Even if domain walls are annihilated before they overclose the universe, their decay products may behave as dangerous relics,
which places additional constraints on the magnitude of the energy bias.
In particular, if domain walls decay into the SM degrees of freedom, the decay products can destroy light elements created at the epoch of BBN,
which conflicts with the standard cosmological scenario.
The ratio between the energy density of domain walls and the entropy density around that time is estimated as
\begin{equation}
\frac{\rho_{\rm wall}}{s}(t) = 2.24 \times 10^{-7}\,\mathrm{GeV}\,\mathcal{A}\left(\frac{g_*(T)}{10}\right)^{3/4} \left(\frac{g_{*s}(T)}{10}\right)^{-1}
\left(\frac{\sigma}{\mathrm{TeV}^3}\right)\left(\frac{t}{1\,\mathrm{sec}}\right)^{1/2}, \label{rho_wall_entropy_ratio}
\end{equation}
where $g_{*s}(T)$ is the relativistic degrees of freedom for the entropy density at the temperature $T$ corresponding to the cosmic time $t$.
According to the constraints on energy injection at the epoch of BBN~\cite{Kawasaki:2004yh,Kawasaki:2004qu},
we must require that the lifetime should be shorter than $t_{\rm ann}\lesssim 0.01\mathrm{sec}$,
if we assume that a significant fraction of the energy density of domain walls is converted into energetic particles.
This condition leads to another lower bound on the magnitude of the energy bias,
\begin{equation}
V_{\rm bias}^{1/4} > 5.07 \times 10^{-4}\,\mathrm{GeV}\,C_{\rm ann}^{1/4}\mathcal{A}^{1/4}\left(\frac{\sigma}{\mathrm{TeV}^3}\right)^{1/4}. \label{BBN_constraint}
\end{equation}
We note that this constraint is derived under the assumption that the decay products strongly interact with SM particles,
and hence it depends on details of underlying particle physics models.

If some stable relics are produced from long-lived domain walls, they would behave as dark matter and
contribute to the energy density of the present universe.
In this case, the tension of domain walls and the magnitude of the energy bias are further constrained from the observed dark matter abundance.
Such a constraint is particularly relevant to axion models, which will be discussed in Sec.~\ref{sec:axion}.

\section{Estimation of gravitational waves from domain walls}
\label{sec3}
\setcounter{equation}{0}
Domain walls having a tension larger than the bound~\eqref{ZKO_bound}
must not exist at the present time, but there is a possibility that they are annihilated before they overclose the universe due to the existence of the energy bias.
It is expected that such collapsing domain walls produce GWs, which are potentially observable today.

The production of GWs from cosmic domain walls was discussed by several authors~\cite{Vilenkin:1981zs,Preskill:1991kd,Chang:1998tb},
while the first quantitative study aiming at comparing the GW signatures and the sensitivities of experiments was
carried out in Ref.~\cite{Gleiser:1998na}. In Ref.~\cite{Gleiser:1998na}, the energy density of the relic GWs
was estimated by solving the evolution of collapsing domain walls numerically and specifying some ansatzes for initial field configurations.
A more improved estimation was performed in Refs.~\cite{Hiramatsu:2010yz,Kawasaki:2011vv}, where the production and the evolution of domain walls in the expanding universe
were investigated based on the field theoretic lattice simulations, and the spectrum of GWs was computed by applying the method introduced in Ref.~\cite{Dufaux:2007pt}.
The results of Refs.~\cite{Hiramatsu:2010yz,Kawasaki:2011vv} were updated in Ref.~\cite{Hiramatsu:2013qaa}
by correcting some error in the numerical code and improving the dynamical range of the simulations.

Let us roughly estimate the energy density of GWs produced by domain walls.
Here we assume that the energy density of domain walls obeys the scaling law~\eqref{rho_wall_scaling}, and that 
the typical time scale of the gravitational radiation is given by the Hubble time $\sim t$.\footnote{It can be shown that the amplitude of GWs
produced by domain walls in the friction dominated regime is much smaller than that produced in the scaling regime, because of their small velocity~\eqref{v_friction_regime}
and curvature radius~\eqref{Rwall_friction_regime}~\cite{Nakayama:2016gxi}.}
According to the quadrupole formula, the power of the gravitational radiation is given by $P\sim G\dddot{Q}_{ij}\dddot{Q}_{ij}\sim M^2_{\rm wall}/t^2$,
where $Q_{ij}\sim M_{\rm wall}t^2$ is the quadrupole moment of domain walls,
and $M_{\rm wall}\sim\sigma\mathcal{A}t^2$ is their mass energy.
Therefore, the energy density of GWs $\rho_{\rm gw}\sim Pt/t^3$ reads
\begin{equation}
\rho_{\rm gw} \sim G\mathcal{A}^2\sigma^2. \label{rho_gw_quadrupole}
\end{equation}
From this estimate we expect that the energy density of GWs produced by domain walls
is proportional to the square of their tension $\sigma^2$ and remains almost constant.

Strictly speaking, the above quadrupole formula cannot be directly applied to domain walls, since it is 
only valid in the far-field regime, while the domain wall network is an spatially extended medium.
In order to check the validity of this estimate, an alternative formalism must be employed.
The formalism to compute the production of GWs from dynamical scalar fields is
descried in Ref.~\cite{Dufaux:2007pt}.
In this approach, the spectrum of GWs can be numerically computed by using Green functions
with transverse-traceless parts of the stress-energy tensor of the scalar field.
In this way, the features shown in Eq.~\eqref{rho_gw_quadrupole} can be checked by performing detailed numerical simulations.

In Refs.~\cite{Hiramatsu:2010yz,Kawasaki:2011vv,Hiramatsu:2013qaa},
numerical simulations of domain walls were performed with the aim of computing the spectrum of GWs produced by them.
In the numerical studies, the evolution of the real scalar field $\phi$ in the simple toy model with $Z_2$ symmetry [Eqs.~\eqref{Lagrangian_Z2} and~\eqref{potential_Z2}] was investigated by solving 
the field equation in the Friedmann-Robertson-Walker background,
\begin{equation}
\ddot{\phi} + 3H\dot{\phi} - \frac{\nabla^2}{R^2(t)}\phi + \frac{dV}{d\phi} = 0,
\end{equation}
where the potential $V(\phi)$ is given by Eq.~\eqref{potential_Z2}.
The simulations were executed in the 3D cubic lattice with the periodic boundary condition.
The radiation dominated background [$R(t)\propto t^{1/2}$] was assumed in Refs.~\cite{Hiramatsu:2010yz,Hiramatsu:2013qaa},
while the evolution in the matter dominated background [$R(t)\propto t^{2/3}$] was also investigated in Ref.~\cite{Kawasaki:2011vv}.
It was confirmed that domain walls enter into the scaling regime [Eq.~\eqref{rho_wall_scaling}] at late times in the simulations.
The effect of the bias term was also investigated in Ref.~\cite{Hiramatsu:2010yz} by adding Eq.~\eqref{bias_potential_Z2} in the scalar potential.
The results of the simulations showed that the collapse of domain walls occurs for sufficiently large $\epsilon$,
where the parameter $\epsilon$ controls the magnitude of the energy bias [Eq.~\eqref{energy_bias_Z2}],
and that the time scale of the collapse agrees with the estimate in Eq.~\eqref{t_ann_general}.

\begin{figure}[htbp]
\begin{center}
\includegraphics[scale=1.0]{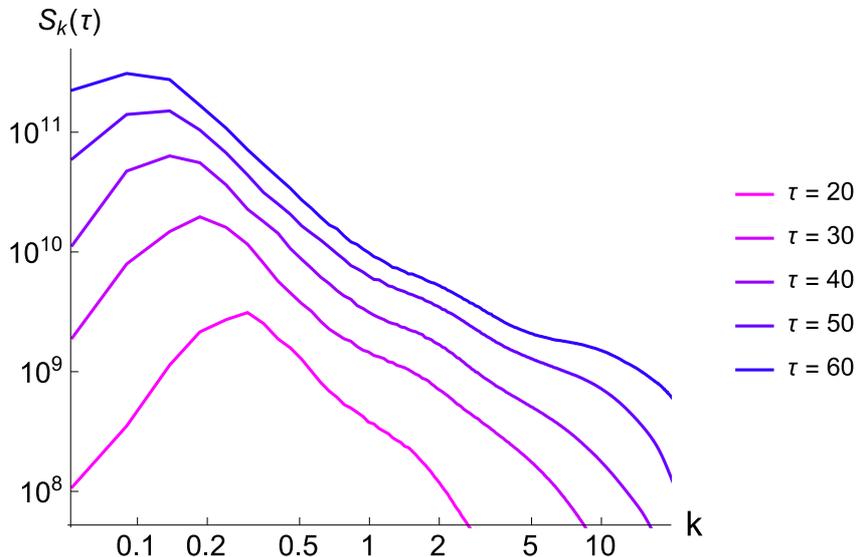}
\end{center}
\caption{The spectrum of GWs [Eq.~\eqref{definition_S_k}] for five different conformal times $\tau=20$, $30$, $40$, $50$, and $60$
obtained in Ref.~\cite{Hiramatsu:2013qaa}.
All dimensionful quantities are shown in the unit of $v=1$, where $v$ is the VEV of the scalar field [see Eq.~\eqref{potential_Z2}].
}
\label{fig:S_k}
\end{figure}

From the configuration of the scalar field in the numerical simulations, one can estimate the spectrum of GWs produced by them.
The spectrum can be computed by using Green functions with transverse-traceless parts of the stress-energy tensor of the scalar field~\cite{Dufaux:2007pt}.
The results obtained in Ref.~\cite{Hiramatsu:2013qaa} are shown in Figure~\ref{fig:S_k}, where $S_k$ represents the spectrum of GWs per unit logarithmic frequency interval,
\begin{equation}
S_k(t) = \frac{2\pi^2 V R^4(t)}{G}\frac{d\rho_{\rm gw}}{d\ln k}(t), \label{definition_S_k}
\end{equation}
with $V$ being the volume of the comoving simulation box.
The spectrum has a peak at the scale corresponding to the Hubble radius.
We note that the horizontal axes in Figure~\ref{fig:S_k} represents the comoving wavenumber, and hence the 
location of the peak $k_{\rm peak}$ shifts according to $k_{\rm peak}/R(t) \sim H(t)$.
Since the smallest scale of the structure of domain walls is given by their core width $\delta$ [see Eq.~\eqref{width_Z2}],
the spectrum falls off at a large wavenumber corresponding to that scale, $k/R(t) \sim \delta^{-1}$.
Furthermore, $S_k$ increases as $\sim k^3$ for $k<k_{\rm peak}$, and decreases as $\sim k^{-1}$ for $k>k_{\rm peak}$.
The behavior $S_k\propto k^3$ at small $k$ can be deduced from causality~\cite{Caprini:2009fx,Hiramatsu:2013qaa}.

The estimate in Eq.~\eqref{rho_gw_quadrupole} can be checked by computing the following quantity
\begin{equation}
\tilde{\epsilon}_{\rm gw} \equiv \frac{1}{G\mathcal{A}^2\sigma^2}\left(\frac{d\rho_{\rm gw}}{d\ln k}\right)_{\rm peak}, \label{def_tilde_epsilon_gw}
\end{equation}
where the subscript ``peak" means that the quantity is evaluated at the peak of $S_k$.
The results of numerical simulations clearly show that the value of $\tilde{\epsilon}_{\rm gw}$ remains almost constant
after domain walls enter into the scaling regime, and it is estimated as~\cite{Hiramatsu:2013qaa}
\begin{equation}
\tilde{\epsilon}_{\rm gw} \simeq 0.7\pm 0.4,
\end{equation}
where the error corresponds to the statistical uncertainty.
Furthermore, the value of $\tilde{\epsilon}_{\rm gw}$ hardly depends on the choice of the value of the parameter $\lambda$, which determines the tension $\sigma$ [see Eq.~\eqref{tension_Z2}].
These facts are consistent with the expectation that the amplitude of GWs is given by Eq.~\eqref{rho_gw_quadrupole} during the scaling regime.

Let us estimate the peak amplitude of GWs produced by long-lived domain walls.
The spectrum of GWs at the cosmic time $t$ is characterized by the following quantity~\cite{Maggiore:1900zz,Maggiore:1999vm}:
\begin{equation}
\Omega_{\rm gw}(t,f) = \frac{1}{\rho_c(t)}\frac{d\rho_{\rm gw}(t)}{d\ln f},
\end{equation}
where $f=k/2\pi R(t)$ is the frequency corresponding to the comoving wavenumber $k$.
From Eq.~\eqref{def_tilde_epsilon_gw}, we have the peak amplitude at the annihilation time of domain walls,
\begin{equation}
\Omega_{\rm gw}(t_{\rm ann})_{\rm peak} = \frac{1}{\rho_c(t_{\rm ann})}\left(\frac{d\rho_{\rm gw}(t_{\rm ann})}{d\ln k}\right)_{\rm peak} = \frac{8\pi\tilde{\epsilon}_{\rm gw}G^2\mathcal{A}^2\sigma^2}{3H^2(t_{\rm ann})}. \label{Omega_gw_t_dec_peak}
\end{equation}
Here, we assume that the production of GWs is suddenly terminated at $t=t_{\rm ann}$,\footnote{This assumption is not rigorous since the collapse of domain walls is not instantaneous,
and they may continue to produce GWs until they completely disappear.
This ambiguity can be incorporated into the definition of $T_{\rm ann}$ or the uncertainty of the parameter $C_{\rm ann}$ in Eq.~\eqref{T_ann_general}.}
and that it happens during the radiation dominated era.
Then, the peak amplitude of GWs at the present time $t_0$ is given by
\begin{align}
\Omega_{\rm gw}h^2(t_0) &= \frac{\rho_{\rm gw}(t_0)h^2}{\rho_c(t_0)} = \frac{\rho_c(t_{\rm ann})h^2}{\rho_c(t_0)}\left(\frac{R(t_{\rm ann})}{R(t_0)}\right)^4\Omega_{\rm gw}(t_{\rm ann}) \nonumber\\
&= \Omega_{\rm rad}h^2\left(\frac{g_*(T_{\rm ann})}{g_{*0}}\right)\left(\frac{g_{*s0}}{g_{*s}(T_{\rm ann})}\right)^{4/3}\Omega_{\rm gw}(t_{\rm ann}), \label{Omega_gw_h2_t0}
\end{align}
where $g_{*0} = 3.36$ and $g_{*s0} = 3.91$ are the effective relativistic degrees of freedom at the present time for the energy density and the entropy density, respectively,
$\Omega_{\rm rad}h^2 = 4.15\times 10^{-5}$ is the density parameter of radiations at the present time, and $h = H_0/100\,\mathrm{km}\cdot\mathrm{sec}^{-1}\mathrm{Mpc}^{-1}$ is the reduced Hubble parameter.
In the first line of Eq.~\eqref{Omega_gw_h2_t0}, we used the fact that the energy density of GWs is diluted as $\rho_{\rm gw}\propto R^{-4}(t)$ for $t>t_{\rm ann}$.
From Eqs.~\eqref{Omega_gw_t_dec_peak} and~\eqref{Omega_gw_h2_t0}, we obtain
\begin{equation}
\Omega_{\rm gw}h^2(t_0)_{\rm peak} = 7.2\times 10^{-18}\,\tilde{\epsilon}_{\rm gw}\mathcal{A}^2\left(\frac{g_{*s}(T_{\rm ann})}{10}\right)^{-4/3}\left(\frac{\sigma}{1\,\mathrm{TeV}^3}\right)^2
\left(\frac{T_{\rm ann}}{10^{-2}\,\mathrm{GeV}}\right)^{-4}. \label{Omega_gw_h2_t0_explicit}
\end{equation}
Note that the large GW amplitude is predicted if the tension $\sigma$ is large and the annihilation temperature $T_{\rm ann}$ is low,
which corresponds to the case where domain walls lived for a long time.

We can also estimate the peak frequency of GWs in terms of the Hubble parameter at the annihilation time of domain walls,
\begin{align}
f_{\rm peak} &\simeq \left(\frac{R(t_{\rm ann})}{R(t_0)}\right)H(t_{\rm ann}) \nonumber\\
&= 1.1\times 10^{-9}\,\mathrm{Hz}\left(\frac{g_{*}(T_{\rm ann})}{10}\right)^{1/2}\left(\frac{g_{*s}(T_{\rm ann})}{10}\right)^{-1/3}\left(\frac{T_{\rm ann}}{10^{-2}\,\mathrm{GeV}}\right). \label{f_peak}
\end{align}
The high annihilation temperature $T_{\rm ann}$ results in the high peak frequency.
Note that there is a cutoff frequency corresponding to the width of domain walls,
\begin{equation}
f_{\delta} \simeq \left(\frac{R(t_{\rm ann})}{R(t_0)}\right)\delta^{-1} = 2.6\times 10^{16}\,\mathrm{Hz}\left(\frac{g_{*s}(T_{\rm ann})}{10}\right)^{-1/3}\left(\frac{T_{\rm ann}}{10^{-2}\,\mathrm{GeV}}\right)^{-1}\left(\frac{\delta^{-1}}{1\,\mathrm{TeV}}\right),
\end{equation}
which is much higher than the peak frequency.
The results of numerical simulations imply that the spectrum of GWs behave as $\Omega_{\rm gw}\propto f^{-1}$ for the intermediate frequency range $f_{\rm peak}<f<f_{\delta}$.

So far we have assumed that the annihilation of domain walls happens during the radiation dominated era.
If it happens before reheating, the above estimates are modified accordingly.
Let us assume that the energy density of the universe is dominated by that of the inflaton, which behaves as non-relativistic matter,
when domain walls are annihilated. Since the Hubble parameter evolves as $H^2\propto R^{-3}(t)$ at that stage, instead of Eq.~\eqref{Omega_gw_h2_t0} we have
\begin{equation}
\Omega_{\rm gw}h^2(t_0) = \Omega_{\rm rad}h^2\left(\frac{g_*(T_{\rm reh})}{g_{*0}}\right)\left(\frac{g_{*s0}}{g_{*s}(T_{\rm reh})}\right)^{4/3}\left(\frac{H_{\rm reh}}{H_{\rm ann}}\right)^{2/3}\Omega_{\rm gw}(t_{\rm ann}),
\end{equation}
where $T_{\rm reh}$ is the reheating temperature, and $H_{\rm reh}$ and $H_{\rm ann}$ are the Hubble parameters at $T=T_{\rm reh}$ and $T=T_{\rm ann}$, respectively.
In the case of the perturbative decay of the inflaton, the Hubble parameter at this stage is given by~\cite{Giudice:2000ex}
\begin{equation}
H = \left[\frac{5\pi^2 g_*^2(T)}{72g_*(T_{\rm reh})}\right]^{1/2}\frac{T^4}{M_{\rm Pl}T_{\rm reh}^2}.
\end{equation}
Using this relation, we obtain
\begin{align}
\Omega_{\rm gw}h^2(t_0) &\simeq 1.2\times 10^{-18}\,\tilde{\epsilon}_{\rm gw}\mathcal{A}^2\left(\frac{g_{*}(T_{\rm reh})}{g_{*}(T_{\rm ann})}\right)^{8/3}\left(\frac{g_{*s}(T_{\rm reh})}{100}\right)^{-4/3} \nonumber\\
&\quad\times \left(\frac{T_{\rm reh}}{10^4\,\mathrm{GeV}}\right)^{20/3}\left(\frac{T_{\rm ann}}{10^5\,\mathrm{GeV}}\right)^{-32/3}\left(\frac{\sigma^{1/3}}{10^9\,\mathrm{GeV}}\right)^6\quad \text{for} \quad T_{\rm ann} > T_{\rm reh}. \label{peak_amplitude_before_reheating}
\end{align}
Furthermore, the peak frequency reads
\begin{align}
f_{\rm peak} &\simeq 8.9\times 10^{-2}\,\mathrm{Hz} \left(\frac{g_{*s}(T_{\rm reh})}{100}\right)^{-1/3}\left(\frac{g_*(T_{\rm ann})}{100}\right)^{1/2}\nonumber\\
&\quad\times\left(\frac{g_*(T_{\rm reh})}{g_*(T_{\rm ann})}\right)^{1/6}\left(\frac{T_{\rm reh}}{10^4\,\mathrm{GeV}}\right)^{-1/3}\left(\frac{T_{\rm ann}}{10^5\,\mathrm{GeV}}\right)^{4/3} \quad \text{for} \quad T_{\rm ann} > T_{\rm reh}. \label{peak_frequency_before_reheating}
\end{align}

The spectrum of GWs for the $Z_N$ symmetric model [Eq.~\eqref{Lagrangian_ZN}] was also analyzed in Ref.~\cite{Hiramatsu:2012sc}.
Similar to the above model with a real scalar field, the spectrum has a peak at the scale corresponding to the Hubble radius,
and the peak amplitude agrees with the estimate based on Eq.~\eqref{rho_gw_quadrupole}.
However, the shape of the spectrum at $f>f_{\rm peak}$ differs from that in the real scalar field model,
and it slightly changes according to the value of $N$.
This $N$ dependence might be caused by the fact that many configurations whose sizes are smaller than the Hubble radius
are produced in the model with large $N$, which results in the enhancement of the amplitude of GWs at high frequencies.

\section{Particle physics models}
\label{sec4}
\setcounter{equation}{0}

In the previous section, we have shown that the amplitude of GWs is determined by two parameters, the tension of the domain wall $\sigma$
and the temperature at the domain wall annihilation $T_{\rm ann}$.
The latter is related to the energy bias $V_{\rm bias}$ for quasi-degenerate vacua [see Eq.~\eqref{T_ann_general}].
The values of $\sigma$ and $V_{\rm bias}$ depend on underlying particle physics models, and hence the prediction for the peak amplitude 
and its frequency differs according to the details of the models.
In this section, we briefly review various particle physics models proposed in the literature
that predict the formation of unstable domain walls and the production of GWs from them.

\subsection{Standard Model Higgs field}
\label{sec:higgs}
Intriguingly, there is a possibility that the dynamics of the SM Higgs field induces the formation of unstable domain walls,
which can produce a significant amount of GWs~\cite{Kitajima:2015nla}.
According to the recent analysis of the effective potential of the Higgs field based on the measured values of the Higgs boson mass and the top quark mass,
our electroweak vacuum is likely to be metastable in the framework of the SM~\cite{Buttazzo:2013uya,Andreassen:2014gha}.
Indeed, the solution of the renormalization group equation for the Higgs self coupling implies that it becomes negative at some energy scale $\Lambda$,
which is much higher than the electroweak scale.
It is probable that there exists some new physics around that scale, which lifts the Higgs potential.
If this is the case, the effective potential of the Higgs field can have two minima, which leads to the formation of Higgs domain walls in the early universe.

The effect of new physics can be modeled by introducing a higher dimensional operator in the Higgs potential,
\begin{equation}
V(\varphi) = \frac{1}{4}\lambda(\varphi)\varphi^4 + \frac{\varphi^6}{\Lambda^2}, \label{V_Higgs}
\end{equation}
where $\varphi$ represents the SM Higgs field value, and $\lambda(\varphi)$ is the field-dependent Higgs self coupling obtained by solving the renormalization group equation
and treating the Higgs field value as the renormalization scale, $\lambda(\mu)=\lambda(\varphi)$.
Here we ignore the quadratic term which leads to the electroweak vacuum, since its effect on the dynamics of the Higgs field is negligible at high energies.
The instability scale $\Lambda$ is sensitive to the top quark mass, and it can take a value from $10^{10}\,\mathrm{GeV}$
to the Planck scale within the error of the measured top quark mass.
Up to the detailed values of $\Lambda$ and the self coupling, the Higgs potential can have two minima, one of which is our electroweak vacuum,
and the other one is at a higher energy scale $\varphi=\varphi_f$ determined by minimizing Eq.~\eqref{V_Higgs}.
Furthermore, one can consider the possibility that the high-scale minimum $\varphi_f$ is just a local minimum, and that two minima are quasi-degenerate, {\it i.e.},
the energy difference $V_{\rm bias}$ between two minima is much smaller than the height of the potential energy $V_0$ separating them (see Figure~\ref{fig:Z2_biased_potential}).

Let us assume that the Higgs potential has quasi-degenerate minima as described above.
If the inflationary scale is sufficiently high, the Higgs field acquires large quantum fluctuations during inflation.
After inflation, the Higgs field takes different values in different patches of the universe, which results in the formation of domain walls.
These domain walls are annihilated when the effect of the energy bias $V_{\rm bias}$ becomes relevant, and subsequently
the electroweak minimum dominates the universe.
The tension of the domain wall can be estimated as~\cite{Kitajima:2015nla}
\begin{equation}
\sigma \sim \left(\frac{\varphi_f^2}{\delta^2}+V_0\right)\delta \sim V_0^{1/2}\varphi_f,
\end{equation}
where the width of the domain wall $\delta \sim \varphi_f/V_0^{1/2}$ can be fixed by minimizing the tension.

The precise values of $\varphi_f$, $V_0$, and $V_{\rm bias}$ should be obtained by solving detailed renormalization group equations,
and they depend on the values of various parameters in the SM such as the top quark mass, the strong gauge coupling, and the Higgs boson mass.
In Ref.~\cite{Kitajima:2015nla}, the magnitude of the energy bias $V_{\rm bias}$ was treated as a free parameter as it can be adjusted by tuning the value of $\Lambda$,
and it was shown that there exists a parameter region in which a significant amount of GWs is produced by long-lived domain walls.
The typical peak frequency reads $f_{\rm peak}\sim 10^{-3}\textendash 10^2\,\mathrm{Hz}$, which is relevant to future direct detection experiments.
The peak frequency cannot be lower than this range, since a smaller value of $\varphi_f$ is required, which cannot be realized in this framework.

The production of GWs from Higgs domain walls was also investigated in Ref.~\cite{Krajewski:2016vbr}.
Contrary to the above discussions, it was concluded that the amplitude of GWs is too small to be observed in the planned detectors.
However, it should be noted that the scenario considered in Ref.~\cite{Krajewski:2016vbr} is different from the above scenario in the sense that
the high-scale minimum is located at a superplanckian value and that two minima are non-degenerate.
In such a case, domain walls do not enter into the scaling regime and collapse soon after the formation, leading to the small amplitude of relic GWs.

\subsection{Axion models}
\label{sec:axion} 

The axion~\cite{Weinberg:1977ma,Wilczek:1977pj} appears in the extensions of the SM with the Peccei-Quinn (PQ) mechanism~\cite{Peccei:1977hh,Peccei:1977ur}, which has been
proposed as a solution to the strong CP problem of quantum chromodynamics (QCD). 
It arises as a pseudo Nambu-Goldstone boson when a hypothetical global U(1) symmetry (called the PQ symmetry) is spontaneously broken.
Its interaction with other particles is suppressed by a large decay constant $F \sim \mathcal{O}(10^9\textendash 10^{11})\,\mathrm{GeV}$,
and hence it is regarded as one of the best motivated candidates of cold dark matter~\cite{Preskill:1982cy,Abbott:1982af,Dine:1982ah}.
Furthermore, string theory suggests the existence of many axion-like particles (ALPs)~\cite{Arvanitaki:2009fg,Cicoli:2012sz}.
For more comprehensive reviews, see Refs.~\cite{Sikivie:2006ni,Ringwald:2012hr,Kawasaki:2013ae,Marsh:2015xka}.

The crucial feature of the axion models is that they predict the formation of domain walls if the PQ symmetry is restored and broken after inflation~\cite{Sikivie:1982qv}.
The global U(1) PQ symmetry is explicitly broken to its subgroup $Z_N$ due to topological charge fluctuations in the QCD vacuum~\cite{DiVecchia:1980yfw,diCortona:2015ldu},
and the effective potential for the axion field $a$ at low energies is described by that in Eq.~\eqref{Lagrangian_ZN},
where the axion mass is given by $m\sim F_{\pi} m_{\pi}/F \sim 6\,\mu\mathrm{eV}\,(10^{12}\,\mathrm{GeV}/F)$, $F_{\pi}\simeq 92\,\mathrm{MeV}$ is the pion decay constant,
$m_{\pi}\simeq 135\,\mathrm{MeV}$ is the pion mass, and $F = v/N$ is the axion decay constant.
In the early universe, first the line-like objects, called global strings, are formed due to the spontaneous breaking of the U(1) PQ symmetry
when the temperature of the universe becomes $T \sim v$. Subsequently, domain walls are formed around the epoch of QCD phase transition.
At that time strings are attached by $N$ domain walls, and the hybrid networks of strings and domain walls, called string-wall systems are formed.\footnote{In some
exceptional cases, domain walls may be formed even if strings do not exist.
For instance, if the initial value of the axion field is tuned to the location which is very close to
the top of the cosine potential~\eqref{Lagrangian_ZN} and its fluctuations are sufficiently large,
domain walls without strings can be formed around the time of QCD phase transition.
Domain walls without strings can also be formed due to the level crossing between the axion and an ALP~\cite{Daido:2015bva,Daido:2015cba},
if there exists an ALP whose mass is comparable to the axion mass around the epoch of QCD phase transition.}
The tension of axionic domain walls is given by Eq.~\eqref{tension_ZN},
\begin{equation}
\sigma \approx 8 m F^2,
\end{equation}
where the approximation implies that there would be some finite corrections in the zero-temperature effective potential~\cite{diCortona:2015ldu,Huang:1985tt},
which we ignore for simplicity.

The evolution of string-wall systems differs according to the number of degenerate minima $N$.
If $N=1$, the systems collapse soon after the formation due to the tension of domain walls~\cite{Vilenkin:1982ks}, and the present energy density of GWs produced from them is too small to observe.
On the other hand, they are stable if $N>1$, and we need to introduce explicit symmetry breaking terms in order to guarantee that they are annihilated before they overclose the universe~\cite{Sikivie:1982qv,Chang:1998tb}.
For instance, Planck-suppressed higher dimensional operators can induce sufficiently small energy bias between $N$ degenerate minima.
It should be noted that such Planck-suppressed operators induce a large CP-violating effect which spoils the original PQ solution to
the strong CP problem~\cite{Ghigna:1992iv,Barr:1992qq,Kamionkowski:1992mf,Holman:1992us,Dine:1992vx,Dobrescu:1996jp},
and that the dimension of those operators must be sufficiently high in order to avoid the experimental limit on the CP violation~\cite{Baker:2006ts}.
These kinds of higher dimensional operators naturally arise
if we assume that the PQ symmetry is an accidental symmetry of an exact discrete symmetry~\cite{Choi:2009jt,Dias:2014osa,Ringwald:2015dsf}.
In Ref.~\cite{Hiramatsu:2010yn}, it was argued that the long-lived domain walls in the axion models with $N>1$ can produce a significant amount of GWs with the peak frequency $f_{\rm peak} \sim 10^{-11}\,\mathrm{Hz}$.
However, it turned out that such a parameter region is excluded since the abundance of cold axions produced by long-lived domain walls
exceeds the observed cold dark matter abundance~\cite{Hiramatsu:2012sc,Kawasaki:2014sqa}.
There still remains some parameter region which avoids all observational constraints, and in such a region the predicted amplitude of GWs is very small, $\Omega_{\rm gw}h^2 \lesssim 10^{-20}$~\cite{Hiramatsu:2012sc}.

A similar argument can be applied to the models with ALPs, but in such models the ALP mass is not necessarily related to its decay constant.
Therefore, one can treat them as two independent parameters in low energy phenomenology. 
In particular, if there exist some couplings between ALPs and SM particles and the ALP mass is sufficiently large,
it is possible to avoid the dark matter overclosure bound, since ALPs produced by long-lived domain walls can decay into radiations.
In Ref.~\cite{Daido:2015gqa}, it was pointed out that domain walls in the ALP models can produce baryon asymmetry of the universe as well as GWs.
A very high peak frequency $f_{\rm peak} \sim \mathcal{O}(100)\,\mathrm{kHz}$ is predicted in this scenario,
since the temperature at the domain wall annihilation must be high, $T_{\rm ann}\gtrsim 10^{11}\,\mathrm{GeV}$, in order to generate sufficiently large baryon asymmetry.

The formation of domain walls and the production of GWs are also predicted in the context of the aligned axion models~\cite{Higaki:2016yqk,Higaki:2016jjh,Farina:2016tgd},
which have been built explicitly by applying the clockwork mechanism discussed in Ref.~\cite{Kaplan:2015fuy}.
In the aligned axion models, the axion $a$ is described in terms of the flat direction of plural axion-like fields $\phi_i$ ($i=1,\dots, N_{\rm ax}$),
where $N_{\rm ax}$ is the total number of the axion-like fields.
A large decay constant $F$ for the axion can be realized even though the actual decay constants $F_i$ for $N_{\rm ax}$ axion-like fields are much smaller than $F$.
Since the symmetry breaking scales $F_i$ are much smaller than the usual PQ scale $F \sim \mathcal{O}(10^9\textendash 10^{11})\,\mathrm{GeV}$,
the CP violating effects from Planck suppressed operators remain small, which naturally explains the high quality of the PQ symmetry.

As an explicit ultraviolet completion of the aligned axion model, one can consider a model based on $N_{\rm ax}$ complex scalar fields associated with $N_{\rm ax}$ global U(1) symmetries.
It is assumed that ($N_{\rm ax} -1$) U(1) symmetries are explicitly broken due to the operators proportional to some small parameters $\epsilon_i \ll 1$.
$(N_{\rm ax} - 1)$ ALPs have masses $m_i\sim \mathcal{O}(\sqrt{\epsilon_i} F_i)$ because of the existence of the explicit symmetry breaking terms, while the other acquires a mass only due to the QCD effect.
In the early universe, $N_{\rm ax}$ U(1) symmetries are spontaneously broken when the temperature of the universe becomes $T \lesssim F_i$, and strings are formed at that epoch.
Subsequently, domain walls with the tension $\sigma = 8 m_i F_i^2$ are formed when the Hubble parameter becomes $H \sim m_i \sim \sqrt{\epsilon_i} F_i$.\footnote{String-wall systems 
may eventually collapse into a single string bundle, since the vacuum may not be disconnected along the direction of the unbroken U(1).
However, such a collapse is not likely to occur if $N_{\rm ax}$ is large and $\epsilon_i$ is sufficiently small, since in this case strings obey the scaling solution, {\it i.e.}, the number of strings per horizon remains $\mathcal{O}(1)$, 
before the formation of domain walls and the size of the hybrid object, which evolves toward the bundle and contains exponentially large number of strings, is far outside the horizon~\cite{Higaki:2016jjh}.}
These domain walls are annihilated around the time of the QCD phase transition, since the potential induced by topological charge fluctuations in the QCD vacuum acts as the energy bias among different domains,
$V_{\rm bias} \sim \Lambda_{\rm QCD}^4$, where $\Lambda_{\rm QCD} \simeq \mathcal{O}(100)\,\mathrm{MeV}$ is the QCD scale.
This fact implies that $T_{\rm ann} \sim 1\,\mathrm{GeV}$, and hence the spectrum of GWs produced by domain walls has a peak at $f_{\rm peak} \sim \mathcal{O}(10^{-7})\,\mathrm{Hz}$.
Based on this fact, it was shown that the present pulser timing observation leads to an upper bound on the fundamental decay constant, $F_i \lesssim \mathcal{O}(100)\,\mathrm{TeV}$~\cite{Higaki:2016jjh}.

\subsection{Supersymmetric models}
\label{sec:susy}

The rich structure of supersymmetric theories gives rise to various possibilities of the formation of domain walls in the early universe.
In Refs.~\cite{Takahashi:2008mu,Dine:2010eb}, the formation of domain walls and the production of GWs
in the context of the spontaneous breaking of discrete $R$ symmetries were discussed.
Let us assume that there exist some hidden SU($N_c$) gauge interactions in addition to those in the SM.
In the supersymmetric extensions of gauge theories, there exist fermionic partners of gauge bosons, called gauginos.
Such gauginos may settle down in a condensate in the early universe due to the corresponding strong gauge forces.
Topological charge fluctuations associated with such gauge forces break the global U(1) $R$ symmetry of the theory down to its discrete subgroup $Z_{2N_c}$,
and this $Z_{2N_c}$ symmetry is spontaneously broken further down to the $Z_2$ subgroup due to the gaugino condensation (see {\it e.g.}~Ref.\cite{Weinberg:2000cr}).
It is known that the effective potential has $N_c$ degenerate vacua after the gaugino condensation, and 
domain walls with the tension 
\begin{equation}
\sigma \sim \Lambda_c^3
\end{equation}
are formed around that time~\cite{Dvali:1996xe,Kovner:1997ca},
where $\Lambda_c$ represents the scale at which the gauge interactions become strong.

In order to avoid the cosmological domain wall problem, it is necessary to introduce a term that induces the energy bias between degenerate vacua.
Such a energy bias is obtained if we assume that there exists a constant term $w_0$ in the superpotential, which explicitly breaks the discrete $Z_{2N_c}$ symmetry.
Note that the magnitude of the constant term should be $w_0 \sim m_{3/2} M_{\rm Pl}^2$ in order to cancel the positive contribution to the cosmological constant
associated with supersymmetry breaking effects, where $m_{3/2}$ is the mass of gravitinos.
This constant term results in the energy bias in the effective potential, $V_{\rm bias} \sim w_0 \Lambda_c^3/M_{\rm Pl}^2\sim m_{3/2} \Lambda_c^3$.
From Eq.~\eqref{t_ann_general}, we see that the annihilation of domain walls occurs when the Hubble parameter becomes comparable to the gravitino mass,
\begin{equation}
H(T_{\rm ann}) \sim \frac{V_{\rm bias}}{\sigma} \sim m_{3/2}.
\end{equation}
This result implies that there is a possibility to probe the gravitino mass from the observation of GWs~\cite{Takahashi:2008mu}.
For instance, if the domain wall annihilation occurs during the radiation dominated era, the peak frequency is given by
$f_{\rm peak}\sim 10^3\,\mathrm{Hz}\,(m_{3/2}/1\,\mathrm{TeV})^{1/2}$.

The formation of domain walls is also predicted in the next-to-minimal supersymmetric SM (NMSSM).
The NMSSM is a possible extension of the minimal supersymmetric SM (MSSM), in which an additional gauge singlet superfield is introduced
in order to provide a solution to the $\mu$-problem~\cite{Kim:1983dt} of the MSSM.
Here, $\mu$ is a dimensionful parameter appearing in the superpotential of the MSSM,
$\mu H_u H_d$ with $H_u$ and $H_d$ being two Higgs doublet superfields, and its magnitude should be of the order of the soft supersymmetry breaking scale
rather than the natural cutoff scale such as the Planck scale.
In the NMSSM, a discrete $Z_3$ symmetry is imposed in order to forbid all dimensionful quantities in the superpotential, and
the $\mu$ term with the appropriate magnitude is induced due to the dynamics of the scalar component $S$ of the singlet superfield, see Refs.~\cite{Maniatis:2009re,Ellwanger:2009dp} for reviews.

The $Z_3$ symmetry is spontaneously broken when the $S$ field acquires expectation values, and domain walls are formed around that time.
The tension of domain walls depends on various parameters including the singlet-Higgs couplings and soft supersymmetry breaking parameters, but typically
$\sigma^{1/3}\sim \mathcal{O}(\mathrm{TeV})$ if the singlet-Higgs couplings are relatively large.
On the other hand, in the decoupling limit where the singlet-Higgs couplings become much smaller than unity, the tension can take much larger values~\cite{Kadota:2015dza},
\begin{equation}
\sigma \sim \kappa \langle S\rangle^3 \gg \mathcal{O}(m_{\rm soft}^3), 
\end{equation}
where $\kappa$ is a dimensionless coupling appearing in the superpotential, $W \supset (1/3)\kappa S^3$,
$\langle S\rangle \sim m_{\rm soft}/\kappa$ is the VEV of the singlet scalar, and $m_{\rm soft}$ represents the soft supersymmetry breaking mass scale.
Due to the large tension, the amplitude of GWs produced from domain walls can be enhanced accordingly.

It was argued that the domain wall problem in the NMSSM
cannot be solved just by introducing Planck-suppressed operators which provide the energy bias among different vacua~\cite{Abel:1995wk},
since such interactions radiatively induce a large tadpole operator that destabilizes the VEV of the singlet field.
One possible solution is to impose various additional symmetries to arrange the form of Planck-suppressed interactions such that the scalar potential only contains a small bias term,
$V_{\rm bias} \sim \zeta m_{\rm soft}^3S+\mathrm{h.c.}$, where $\zeta$ is a loop suppression factor~\cite{Panagiotakopoulos:1998yw}.
Another possibility is to assume that the $Z_3$ symmetry is anomalous for QCD or some hidden strong gauge interactions~\cite{Hamaguchi:2011nm}.
In this case, the energy bias is given by $V_{\rm bias} \sim \Lambda_s^4$, where $\Lambda_s$ is the scale at which the corresponding gauge interactions become strong.
In any case domain walls must be annihilated before the epoch of BBN, since their decay products would destroy light elements [see Eq.~\eqref{BBN_constraint}].
This fact implies that the peak frequency of GWs produced from domain walls should be higher than $f_{\rm peak} \sim 10^{-9}\,\mathrm{Hz}$.
Assuming that the domain wall annihilation occurs just before BBN, one can constrain the parameters of the NMSSM in the decoupling limit from pulsar timing observations~\cite{Kadota:2015dza}.

In Ref.~\cite{Moroi:2011be}, the production of GWs from domain walls formed after thermal inflation was discussed.
Thermal inflation~\cite{Lyth:1995hj,Lyth:1995ka} is introduced in order to suppress the abundance of harmful light long-lived scalar fields, called moduli,
appearing in the context of string cosmology~\cite{Banks:1993en,deCarlos:1993wie}.
In this model, a short period of inflation is driven by the potential energy of a scalar field called flaton, which is trapped at the origin of the scalar potential because of thermal effects.
A discrete $Z_n$ symmetry is imposed in order to guarantee the flatness of the flaton potential, where $n$ is an integer satisfying $n\ge 4$.
After thermal inflation, the thermal effects become irrelevant and the flaton acquires non-zero VEVs, which break $Z_n$ symmetry and lead to the formation of domain walls.
The domain walls can be annihilated if we introduce an additional term that explicitly breaks $Z_n$ symmetry~\cite{Asaka:1997rv,Asaka:1999xd}
or if we assume that the $Z_n$ symmetry is anomalous for QCD~\cite{Preskill:1991kd,Moroi:2011be}.
It was shown that such domain walls can produce a significant amount of GWs if their lifetime is sufficiently long,
and the peak frequency typically lies in the range of $10^{-6}\textendash 10^{-3}\,\mathrm{Hz}$~\cite{Moroi:2011be}.

\section{Implications for present and future observations}
\label{sec5}
\setcounter{equation}{0}

In this section, we discuss implications of GWs from domain walls for ongoing and planned experimental searches.
Here we marginalize the model-dependences described in the previous section, and 
treat $\sigma$ and $T_{\rm ann}$ as free parameters to clarify the parameter region that is relevant to present and future observations.

The leading ground-based interferometer is Advanced LIGO~\cite{TheLIGOScientific:2014jea}, whose first observing run placed limits on the amplitude of
the stochastic GW background $\Omega_{\rm gw} < 1.7\times 10^{-7}$ with 95\% confidences for $20\textendash 86\,\mathrm{Hz}$ by assuming a flat GW spectrum~\cite{TheLIGOScientific:2016dpb},
and these limits are about 33 times tighter than the previous limits set by Initial LIGO and Virgo~\cite{Aasi:2014zwg}.
In addition to them, ground-based interferometer KAGRA~\cite{Aso:2013eba} will soon start to run in Japan.
In Europe, the more advanced ground-based observatory, Einstein Telescope (ET)~\cite{Punturo:2010zz}, is planned 
with the aim of achieving further improvement in sensitivity.
Space-borne interferometers such as eLISA~\cite{Klein:2015hvg,Caprini:2015zlo} and DECIGO~\cite{Kawamura:2006up} are planned to be launched in the future,
and they will enable us to explore lower frequency ranges, which cannot be probed in the ground-based experiments.
Much lower frequencies $\sim 10^{-9}\textendash 10^{-8}\,\mathrm{Hz}$ are probed by using the pulsar timing array (PTA).
Recently, European Pulsar Timing Array (EPTA) set a limit on the amplitude of a flat stochastic GW background $\Omega_{\rm gw}h^2 < 1.2\times 10^{-9}$
at a reference frequency of $f = 1\mathrm{yr}^{-1}$~\cite{Lentati:2015qwp}, and it is about
one order of magnitude tighter than that obtained in previous PTA searches~\cite{Jenet:2006sv,vanHaasteren:2011ni,Demorest:2012bv}.
The sensitivity will be improved in future PTA projects such as SKA~\cite{Janssen:2014dka} and FAST~\cite{Nan:2011um}.

\begin{figure*}[htbp]
\begin{center}
\includegraphics[scale=0.9]{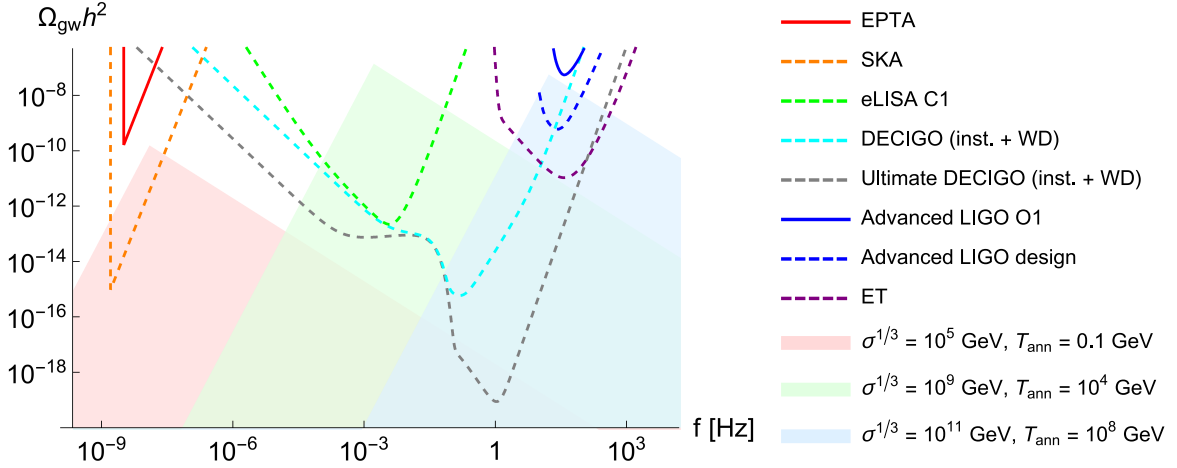}
\end{center}
\caption{The schematics of the sensitivities of present/future GW experiments and GW signatures from domain walls.
Solid lines represent the present upper limits on the GW background obtained by EPTA (red) and Advanced LIGO O1 (blue).
Dashed lines represent the sensitivities of future experiments including SKA (orange), eLISA (green), DECIGO (cyan),
Ultimate DECIGO (gray), Advanced LIGO design (blue), and ET (purple).
The sensitivity curves for DECIGO and Ultimate DECIGO contain both the instrumental noise and the WD confusion noise.
Light colored regions represent typical spectra of GWs from domain walls for
$\sigma^{1/3}=10^5\,\mathrm{GeV}$ and $T_{\rm ann}=0.1\,\mathrm{GeV}$ (light red),
$\sigma^{1/3}=10^9\,\mathrm{GeV}$ and $T_{\rm ann}=10^4\,\mathrm{GeV}$ (light green), and
$\sigma^{1/3}=10^{11}\,\mathrm{GeV}$ and $T_{\rm ann}=10^8\,\mathrm{GeV}$ (light blue).}
\label{fig:sensitivity1}
\end{figure*}

Sensitivities of various ongoing and planned experiments are summarized in Figure~\ref{fig:sensitivity1}.
For the sensitivities of Advanced LIGO, we plot the constraint on arbitrary power spectra (``Advanced LIGO O1")
and the design sensitivity with the assumption of two years observations in Advanced LIGO and Virgo (``Advanced LIGO design")
reported in Ref.~\cite{TheLIGOScientific:2016dpb}.
These lines imply that GW signals above (below) them correspond to $\mathrm{SNR}\ge 2$ ($\mathrm{SNR}\le 2$).
For other interferometers, we assume one year cross-correlation searches and plot the lines with $\mathrm{SNR} = 2$.
The sensitivity curve for ET is produced by using a fitting function in~\cite{ETsensitivity}.
The sensitivity of eLISA depends on the detailed detector configurations, and here we assume the C1 configuration, whose parameters are specified in Refs.~\cite{Klein:2015hvg,Caprini:2015zlo}.
For instrumental noises of DECIGO and Ultimate DECIGO, we used the parameters specified in Ref.~\cite{Kuroyanagi:2014qza}.
It should be noted that GWs produced from white-dwarf (WD) binaries may lead to a significant confusion noise,
which decreases sensitivities at lower frequencies of $f\lesssim 0.1\,\mathrm{Hz}$.
In Figure~\ref{fig:sensitivity1}, we adopt the fitting formula for the WD confusion noise specified in Refs.~\cite{Barack:2003fp,Nishizawa:2011eq}
in addition to instrumental noises of DECIGO and Ultimate DECIGO.
The sensitivities of EPTA and SKA are taken from~\cite{GWplotter,Moore:2014lga}.

In Figure~\ref{fig:sensitivity1}, we also plot the GW signatures from cosmic domain walls for three choices of parameters.
In these plots, we used Eqs.~\eqref{Omega_gw_h2_t0_explicit} and~\eqref{f_peak} to estimate the peak amplitude and frequency.
The spectra are extrapolated based on the frequency dependences implied by the results of numerical simulations,
$\Omega_{\rm gw}\propto f^3$ for $f<f_{\rm peak}$ and $\Omega_{\rm gw}\propto f^{-1}$ for $f>f_{\rm peak}$.
We see that sufficiently large GW signatures are predicted according to the values of $\sigma$ and $T_{\rm ann}$.

\begin{figure*}[htbp]
\begin{center}
\includegraphics[scale=0.85]{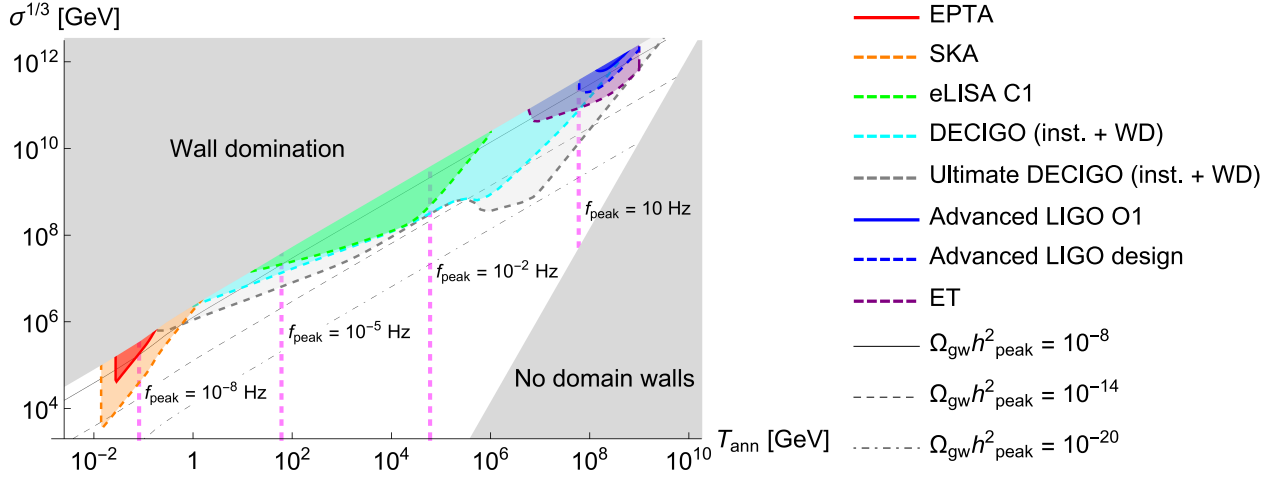}
\end{center}
\caption{Sensitivities of present/future GW experiments in the parameter space of $T_{\rm ann}$ and $\sigma^{1/3}$.
In the colored regions, the peak amplitude of GWs from domain walls estimated based on Eq.~\eqref{Omega_gw_h2_t0_explicit}
exceeds the sensitivity curves plotted in Figure~\ref{fig:sensitivity1}.
Gray regions correspond to the parameter space in which domain walls overclose the universe [not satisfying Eq.~\eqref{wall_domination_constraint}]
or they are not formed since the energy bias is too large [not satisfying Eq.~\eqref{bias_constraint} with $V_0=\sigma^{4/3}$].
The thin black lines represent the contours for the peak amplitude of GWs from domain walls, $\Omega_{\rm gw}h^2_{\rm peak}=10^{-8}$ (solid),
$10^{-14}$ (dashed), and $10^{-20}$ (dot-dashed).
The dashed magenta lines represent the contours for the peak frequency with $f_{\rm peak}=10^{-8}\,\mathrm{Hz}$, $10^{-5}\,\mathrm{Hz}$,
$10^{-2}\,\mathrm{Hz}$, and $10\,\mathrm{Hz}$.}
\label{fig:sensitivity2}
\end{figure*}

Following Ref.~\cite{Nakayama:2016gxi}, in Figure~\ref{fig:sensitivity2} we specify the parameter region of
$T_{\rm ann}$ and $\sigma^{1/3}$ relevant to observations.
The colored regions in Figure~\ref{fig:sensitivity2} correspond to the parameter values 
for which the peak amplitude of GWs from domain walls [Eq.~\eqref{Omega_gw_h2_t0_explicit}]
exceeds the sensitivity curves plotted in Figure~\ref{fig:sensitivity1}.
In Figure~\ref{fig:sensitivity2}, we also plot the parameter region denoted by ``Wall domination", which corresponds to the potential uncertainties 
since the energy density of domain walls dominates the total energy density of the universe [see Eq.~\eqref{wall_domination_constraint}].
Furthermore, the large scale domain walls cannot be formed if the condition shown in Eq.~\eqref{bound_bias} is not satisfied.
Combining Eqs.~\eqref{bound_bias} and~\eqref{T_ann_general}, we obtain the following condition,
\begin{equation}
T_{\rm ann} < 3.04\times 10^4\,\mathrm{GeV}\,C_{\rm ann}^{-1/2}\mathcal{A}^{-1/2}\left(\frac{g_*(T_{\rm ann})}{10}\right)^{-1/4}\left(\frac{\sigma^{1/3}}{\mathrm{GeV}}\right)^{-3/2}\left(\frac{V_0}{\mathrm{GeV}^4}\right)^{1/2}. \label{bias_constraint}
\end{equation}
Up to the value of $V_0$, this condition gives an upper limit on $T_{\rm ann}$.
Here we take $V_0 = \sigma^{4/3}$ as a typical estimate of the height of the potential barrier.\footnote{From Eq.~\eqref{typical_estimate_sigma}, we see that the choice $V_0 = \sigma^{4/3}$
corresponds to $\delta^{-1} \sim \sigma^{1/3}$, which is satisfied in the toy model given by Eqs.~\eqref{Lagrangian_Z2} and~\eqref{potential_Z2} if $\lambda\simeq \mathcal{O}(1)$.
We also note that the condition~\eqref{bound_bias} is always satisfied for domain walls in axion models [Eq.~\eqref{Lagrangian_ZN}],
since Eqs.~\eqref{width_ZN},~\eqref{typical_estimate_sigma},~\eqref{bound_bias}, and~\eqref{t_ann_general} imply $m \gtrsim H_{\rm ann}$,
which holds after the formation of domain walls.}
The corresponding parameter region is denoted by ``No domain walls" and shown in Figure~\ref{fig:sensitivity2}.

Both in Figure~\ref{fig:sensitivity1} and in Figure~\ref{fig:sensitivity2}, we have assumed that
the annihilation of domain walls occurs during the radiation dominated era.
If it occurs before reheating, the amplitude and frequency of GWs are modified
according to the value of the reheating temperature [see Eqs.~\eqref{peak_amplitude_before_reheating} and~\eqref{peak_frequency_before_reheating}]. 

From Figure~\ref{fig:sensitivity2}, we see that the significantly large GW amplitude is predicted 
if the energy density of domain walls is close to dominate the total energy density of the universe.
It should be noted that recent observational results by EPTA and Advanced LIGO already exclude some parameter spaces.
Future observations with improved sensitivities are expected to probe much wider ranges of the parameter space,
and from such observations we will obtain richer information about high energy physics beyond the SM.

\section{Conclusion}
\label{sec6}
Various well-motivated particle physics models
predict the formation of unstable domain walls in the early universe,
and it is possible to probe such models by observing GWs produced by them.
The signatures of GWs can be characterized by two quantities, the tension of domain walls $\sigma$ and the
temperature at the annihilation of them $T_{\rm ann}$.
Values of these parameters depend on the details of models, and they range over many orders of magnitude.
Accordingly, future broadband observations of GWs including PTA, ground-based, and space-borne interferometers
will allow us to explore new physics at various energy scales, some of which cannot be reached in the conventional laboratory experiments.
Assuming that the annihilation of domain walls occurs during the radiation dominated era, we have shown that
the ranges of $10^{-2}\,\mathrm{GeV}\lesssim T_{\rm ann} \lesssim 10^9\,\mathrm{GeV}$ and
$10^3\,\mathrm{GeV}\lesssim\sigma^{1/3}\lesssim 10^{12}\,\mathrm{GeV}$ can be covered by future experiments.

So far we have estimated the spectrum of GWs from domain walls based on a naive extrapolation of the results obtained
in the field theoretic lattice simulations.
However, it is difficult to estimate the spectrum of GWs accurately over broad frequency ranges 
due to the limitation of the dynamical ranges of the simulations.
In order to resolve this difficulty, it will be necessary to develop 
some alternative method to compute the spectrum of GWs analytically.
Such an approach would enable us to estimate the signatures of GWs more quantitatively,
and it can be used to distinguish the signal of domain walls from that of other sources.
Furthermore, the results of numerical simulations of domain walls associated with $Z_N$ symmetric models
imply that the shape of the spectrum slightly differs depending on the value of $N$~\cite{Hiramatsu:2012sc}.
The $N$-dependent feature in the spectrum of GWs might be regarded as an additional information to distinguish different models,
and it deserves further investigation.

The search for cosmological GWs would have a great impact on high energy physics and cosmology.
The collapse of domain walls will provide a possible way to interpret the results of forthcoming GW experiments.
Even if there is no evidence of a signal, such information can be used to constrain various particle physics models beyond the SM.

\section*{Acknowledgments}
The author would like to thank Takashi Hiramatsu, Kenji Kadota, Masahiro Kawasaki, and Toyokazu Sekiguchi
for the collaborations on the topics in this review.


\end{document}